\documentclass{aa} 
\usepackage{graphicx}
\usepackage{txfonts}

\begin{document}

\title{The flow field in the sunspot canopy }

\author{R. Rezaei\inst{1}, R. Schlichenmaier\inst{1}, C.A.R. Beck\inst{1}, and L.R. Bellot Rubio\inst{1,2}}

\institute{Kiepenheuer-Institut f\"ur Sonnenphysik, Sch\"oneckstr. 6, 79104 Freiburg, Germany
\and
Instituto de Astrof\'isica de Andaluc\'ia (CSIC), Apdo. 3004, 18080 Granada, Spain }

\date{Received 10 March 2006/Accepted 28 March 2006}

\titlerunning{The flow field at the sunspot canopy}
\authorrunning{Rezaei et al.}

\abstract
{}
{We investigate the flow field in the sunspot canopy using 
simultaneous Stokes vector spectropolarimetry of 
three sunspots ($\theta$\,=\,27\degr, 50\degr, 75\degr) and 
their surroundings in visible (630.15 and 630.25\,nm) 
and near infrared (1564.8 and 1565.2\,nm) neutral iron lines.}
{To calibrate the Doppler shifts, we compare 
an absolute velocity calibration using 
the telluric $\textrm{O}_2$--line at 630.20\,nm and a relative velocity calibration 
using the Doppler shift of Stokes $V$ profiles in the umbra under the 
assumption that the umbra is at rest. 
Both methods yield the same result within the calibration 
uncertainties ($\sim$\,150 m\,s$^{-1}$). 
We study the radial dependence of  Stokes $V$ profiles 
in the directions of disk center and limb side. }
{Maps of Stokes $V$ profile shifts, polarity, amplitude asymmetry, field strength  
and magnetic field azimuth provide strong evidence for the presence of a magnetic canopy 
and for the existence of a radial outflow in the canopy.}{  
Our findings indicate that the Evershed flow does not cease abruptly at 
the white--light spot boundary, but that at least a part of the penumbral Evershed flow 
continues into the magnetic canopy. 
}
\maketitle
\normalsize
\section{Introduction}

From the pioneering work of Evershed (1909), we know that there is a flow field  
in the sunspot penumbra, which leads to a line shift and line asymmetry. 
For sunspots outside the disk center, it manifests in blue shifted spectral lines in 
the center side and red shifted spectral lines in the limb side. 
The penumbral fine structure has a close relation with the Evershed
flow. It is generally accepted that the flow channels are more horizontal
than the background field (Solanki 2003). 
Because spectropolarimetric data has lower 
spatial resolution than narrow band magnetograms, many questions regarding both 
the Evershed flow and the fine structure of the penumbra have 
remained (e.g. Solanki 2003; Bellot Rubio 2004). 
There are also controversial arguments about the continuation of 
the Evershed flow outside the sunspot.

Brekke \& Maltby (1963) studied horizontal variations of the Evershed flow. 
Investigating 2500 sunspot spectra, they reported that at the outer 
penumbral boundary ``the velocity falls abruptly to zero''. 
Wiehr et al. (1986) observed a sharp decrease of the Evershed effect and magnetic field 
at the visible boundary of sunspots. Using only Stokes $I$ spectra, they argued that 
line--core shifts and line asymmetries are ``strongly'' limited to the continuum 
boundary of sunspots. 
Wiehr \& Balthasar (1989), Schr\"oter et al. (1989) and Title et al. (1993) confirmed 
the result of Brekke \& Maltby (1963). Wiehr (1996) renewed his argument 
that the Stokes $I$ profile asymmetries of Ni\,I 543.6\,nm (g\,=\,0.5) and 
Fe\,I 543.5\,nm (g\,=\,0) ``disappear'' within less than one arcsec from the penumbral border. 
Hirzberger \& Kneer (2001) observed two sunspots at 
heliocentric angles of 31\degr and 20\degr; they reported a sharp decrease of 
the Evershed flow (intensity profile asymmetry) at the penumbral 
boundary, using Stokes $I$ of the non--magnetic 
Fe\,I 557.6\,nm and Fe\,I 709.0\,nm lines. 

In contrast, 
Sheeley (1972) stated that there is a horizontal flow in the plage--free 
photosphere surrounding sunspots with an average 
velocity of $\sim$\,0.5\,--\,1\,km\,s$^{-1}$. 
K\"uveler \& Wiehr (1985)  did not find a sharp change in the Evershed flow at the penumbral boundary. 
Dialetis et al. (1985), Alissandrakis et al. (1988), and  B\"orner \&  Kneer (1992) 
confirmed this result. 
Giovanelli \& Jones (1982) reported magnetogram observations of diffuse, almost 
horizontal magnetic field surrounding two spots at moderate heliocentric angles. 
Taking Stokes $V$ and $I$ profiles of the 1.56 $\mu$m iron lines for two limb spots, 
Solanki, Montavon, \& Livingston (1994) found that the magnetic field of sunspots continues beyond 
the visible boundary and forms  an extensive canopy above a non--magnetic layer.  
These authors computed the canopy base height and reported that around 10\,\% 
of the Evershed flow  continues into the magnetic canopy. 
Rimmele (1995a, b) also found no sharp boundary in time--averaged velocity maps of a sunspot 
close to the disk center. After observing the Fe\,I 557.6\,nm 
line with a narrow--band filtergraph, he 
obtained different velocities at different bisector levels which do not  decrease 
abruptly at the sunspot boundary. 
Solanki et al. (1999) repeated their claims about continuation of the Evershed flow 
by considering $V$ and $I$ signals of the 1.56 $\mu$m iron lines for another sunspot close to 
the limb ($\mu\,=\,\cos\theta$\,=\,0.22). 
Computing the amplitude of the azimuthal velocity variation, Schlichenmaier \& Schmidt (2000) 
did not find a drop of the line--core velocity of Fe\,II 542.5\,nm at the penumbral 
boundary. 
Therefore, this long standing disagreement about continuation/termination 
of the flow field at the sunspot boundary was intensified.

However, it is important to note that none of the mentioned authors observed 
the full Stokes parameters of the target spots. 
Moreover, a majority of them only used Stokes $I$ profiles, which suffer from  
stray light contamination. Here, we present simultaneous 
spatially co--aligned full Stokes spectropolarimetric observations of three 
sunspots and their surroundings in visible (630.15 and 630.25\,nm) and 
near infrared (1564.8 and 1565.2\,nm) neutral iron lines. 
These co--temporal and co--spatial observations of the full Stokes vector 
provide valuable information not only about magnetic field strength in the canopy, 
but also about the field orientation as emphasized by Solanki et al. (1994).
The near--IR neutral iron lines mostly form deep in the 
atmosphere, while the contribution function of the visible 
iron lines at 630\,nm peaks in higher layers (Cabrera Solana et al. 2005). 
As these lines form in different atmospheric layers and have large Zeeman sensitivities, 
they provide a powerful tool to study the properties of the canopy, 
its flow field and vertical structure.

Considering the fact that the Stokes profiles of $Q$, $U$, and $V$  are 
formed only in a magnetized atmosphere, 
this data set contains information regarding  the extension of the magnetic 
canopy and the Evershed flow outside the white--light sunspot boundary. 
In sections 2 and 3, we explain our observations and data reduction in detail. 
In section 4, the spatial variation of the flow field is investigated using 
two independent methods. Conclusions and comparisons are discussed in Sect. 5.

\section{Observations}

Three isolated sunspots (cf.~Fig.~1) were observed at the German 
Vacuum Tower Telescope (VTT) in Tenerife, August 2003. 
Simultaneous co--aligned spectropolarimetric data with the Tenerife Infrared 
Polarimeter (TIP) and the POlarimetric LIttrow Spectrograph (POLIS) were recorded. 
The TIP (Collados 1999; Mart\'inez Pillet et al. 1999) 
observed full Stokes profiles of the 
infrared iron lines at 1564.8\,nm\,(g\,=\,3) and 1565.2\,nm\,(g\,=\,1.53).
The visible neutral iron lines at 630.15\,nm, 630.25\,nm, and 
Ti 630.38\,nm were observed with the POLIS (Schmidt et al. 2003; Beck et al. 2005a).

For the spots  1 and  3, the scanning step--size was 0.35 arcsec, while for spot  2 
it was 0.4 arcsec. Spatial sampling of the TIP maps along the slit was 0.35 arcsec. 
For the POLIS, the spatial sampling 
along the slit was 0.15 arcsec (Beck et al. 2005a). To improve the signal--to--noise ratio 
and to have a comparable spatial resolution as in the infrared data, we 
 bin the data along the slit by a factor of 3. 
The pixel size of these new maps along the slit is 0.44 arcsec. 
The spectral sampling of 2.97 pm for TIP 
and 1.49 pm for POLIS leads to a velocity dispersion of 
570 m\,s$^{-1}$ and 700 m\,s$^{-1}$ per pixel, respectively. 
Seeing conditions during the observations were good and stable. 
We used the Correlation Tracker System (Schmidt \& Kentischer 1995; Ballesteros et al. 1996) 
to compensate for image motion. 
We estimate the spatial resolution to be about 1.0 arcsec for the spots  1 and  2 
and 1.5 arcsec for the spot  3. 
The duration of the observations for each sunspot was around 10 minutes.

\begin{table}
\begin{center}
\caption{Properties of the observed sunspots. $\theta$ is the heliocentric angle. 
The last column is the scanning step--size in arcsec.}
\begin{tabular}{c c c c c} 
\hline
Spot No. & Date  & Active Region & $\theta$ (\degr) & step--size\\ \hline
1   & 09/08/2003 & 10430  & 27  & 0.35\\
2   & 03/08/2003 & 10425  & 50  & 0.40\\
3   & 08/08/2003 & 10421  & 75  & 0.35\\
\hline
\end{tabular}
\end{center}
\end{table}

The spectropolarimetric data of both TIP and 
POLIS have been corrected for instrumental effects and telescope 
polarization with the procedures described in Collados (1999; 
Mart\'inez Pillet et al. 1999)   
for the TIP and Beck et al.(2005a, b) for the POLIS. 
Remaining cross--talk in our data sets is of the order of $10^{-3}$\,$I_c$. 
Table 1 summarizes the characteristics of the three 
observations. The spot closest to the limb was at $\theta$\,=\,75 deg.

\begin{figure*}
\centerline{\resizebox{18cm}{!}{\includegraphics{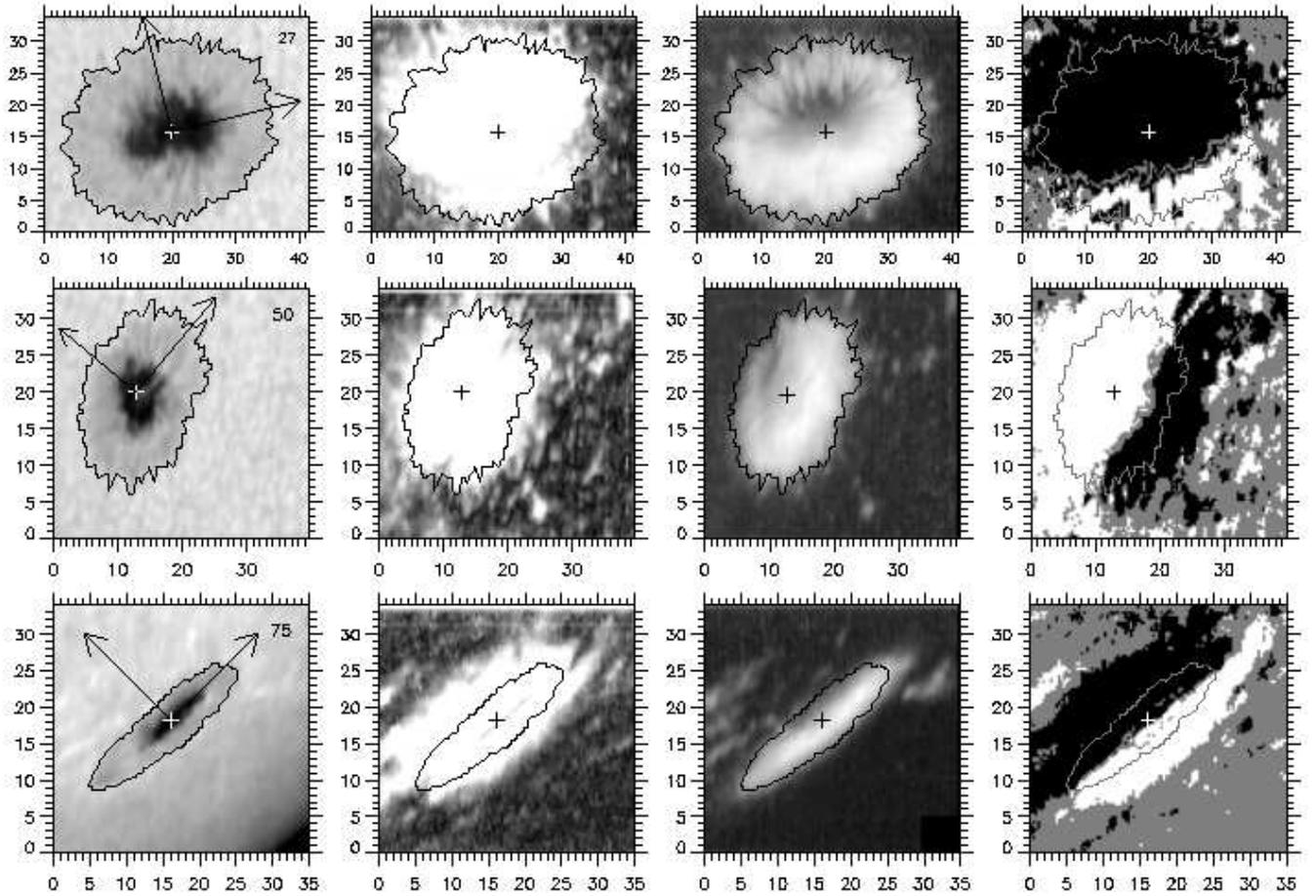}}}

\caption[]{{\em From left to right:} continuum intensity maps  of the sunspots,  
integrated circular polarization ($\int |V(\lambda)| d\lambda$), 
the integrated linear polarization signal ($L_t\,=\,\int L(\lambda) d\lambda$), 
and polarity in Fe\,I 1564.8\,nm line. 
{\em From top to bottom:} spot  1, spot  2, and spot  3. 
Axis labels are spatial dimensions in arcsec. 
The numbers on the upper--right corners are the heliocentric angles of the spots.
In the white--light maps, the left arrows show the disk center directions. 
Note that the magnetic field clearly extends over the white--light boundary along the line 
connecting sunspot to the disk center in the $V$ maps and perpendicular to this line in $L_t$ maps.} 
\end{figure*}


\section{Data analysis}
In this section, we explain the data reduction and present our definition for 
the sunspot boundary. Then we investigate two independent methods of velocity 
calibration. Finally, we extract vector magnetic field parameters.

\subsection{Pre-processing}

We use a continuum intensity threshold to define umbral data points and their mean position, 
the umbral center. 
After that, we draw a radial line from the umbral center in azimuthal directions (in 0.5\degr \ steps). 
The point at which this line crosses the penumbral white--light boundary defines the 
sunspot border and radius in  this direction. 
The continuum threshold which are used at the 
penumbral border are 0.8 and 0.9 $I_c$ for the visible and infrared data respectively. 
These crossing points define a closed path (contour) around each 
sunspot (cf. Fig. 1, left column). 

All polarization signals, $Q(\lambda)$, $U(\lambda)$, $V(\lambda)$, and 
the total linear polarization profiles, $L(\lambda)$\,=\,$\sqrt{Q(\lambda)^2 + U(\lambda)^2}$, are 
normalized by the local continuum intensity, $I_c$, for each pixel, i.e., 
$V(\lambda)$\,$\equiv$\,$V(\lambda)/I_c$. 
The rms noise level of Stokes parameters in the continuum  
is $\sigma$\,=\,3\,--\,5\,$\times 10^{-4}$\,$I_c$ for the 
infrared lines and $\sigma$\,=\,1.5\,$\times 10^{-3}$\,$I_c$ for the visible lines.  
Only pixels with $V$ signals greater than 5\,$\sigma$ for TIP (3\,$\sigma$ for POLIS) are 
included in the analysis. Positions and amplitudes of the profile extrema  
in all Stokes parameters are obtained by fitting a parabola to each lobe.

For the POLIS data, we apply a correction along the slit before calibrating the velocity. 
In this direction, there is a curvature 
in the profiles due to the small focal length of the spectrograph. 
For each scan step, a third order polynomial is fitted to positions of the 630.20\,nm telluric line--cores  
along the slit. 
All profiles in that scan step are shifted with the resulting curve. 
This correction amounts at most to $\sim$\,4 (spectral) pixels shift between upper most and lower most 
spatial pixels. 
Final maps of the line--core velocity of the telluric line are quite uniform with a small 
dispersion (standard deviation) in the umbral region, 
e.g. around 0.11\,pixel ($\equiv$\,77 m\,s$^{-1}$) for the spot 1. 
Using this telluric rest--frame, we calibrate  Stokes $V$ and $I$ shifts.

\subsection{Velocity calibration}

We consider two different calibration methods. The first method is 
based on Stokes $V$ profiles of the umbra. Assuming that the umbra is at rest, 
we select quite antisymmetric Stokes $V$ profiles. 
The Stokes $V$ profile is expected to be strictly antisymmetric with respect to 
its zero--crossing wavelength, if no velocity gradients with height is present and if LTE applies 
(Auer \& Heasley 1978; Landi Degl'Innocenti \& Landolfi 1983).
Hence, the mean position of these profiles  
is assumed to define a {\it rest--frame}. 
To confirm the assumption of the umbra at rest, we perform an absolute velocity calibration
using the telluric lines in the POLIS data. 
At the end of this section, we compare the two calibration methods. 
Results are completely consistent and differences are less than the calibration error.

\subsubsection{Velocity calibration using Stokes $V$ profiles of the umbra}
We attribute a relative amplitude asymmetry, $\delta a$, to each $V(\lambda)$ profile. 
It is defined as the  
asymmetry between the amplitudes of the two lobes of Stokes $V$ by: 
$$ \delta a =  \frac{a_{b} - a_{r}}{a_{b} + a_{r}},$$
where $a_{b}$ and $a_{r}$ are the absolute amplitudes of 
blue and red lobes, respectively. 
The center of each $V(\lambda)$ profile is defined as the midway between its maximum and minimum. 
This quantity is better than deriving the zero--crossing by a linear regression, because 
it is less affected by magneto--optical effects. 
First, all umbral profiles with an amplitude asymmetry lower than a threshold are 
selected, which defines a class corresponding to a row in Table 2. Then, the relative velocity of 
each selected profile with respect to the mean value of the class (reference point) is determined. 
So each class has a velocity dispersion around its mean value which is defined to be zero.  
The velocity dispersions for different classes are similar ($d_1$ and $d_2$ in Table 2).
These are the main uncertainties in the Stokes $V$ velocity calibration. 
The difference between reference point of each class and the $|\delta a| \le 0.05$ class is given 
in Table 2 (fourth and last columns). 
We select the mean position of a set of quite antisymmetric umbral 
profiles  with $|\delta a| \le 0.05$ as the sunspot rest--frame, which is a compromise between low 
amplitude asymmetry and good statistics (Table 2). 

\begin{table}
\begin{center}
\caption{{\em From top to bottom:} spot 1, 2, and 3.
The first column is the maximum amplitude asymmetry in each class. 
$d_1$ ($d_2$) shows the velocity dispersion of the selected umbral profiles in 
infrared (visible) lines in each row. 
\% IR (vis) gives the number of umbral profiles in percent of all umbral pixels with 
$\delta a$ less than the threshold.  
The fourth and the last columns are difference of velocity reference point of each class  
relative to the final result ($|\delta a| \le$ 0.05). 
The last row gives number of total umbral profiles. 
Velocities are in m\,s$^{-1}$.}
\begin{tabular}{c c c c c c c} 
\hline
max. $|\delta a|$ & $d_1$ & \% IR & ref: IR &  $d_2$ & \% vis & ref: vis\\
\hline
0.10  & 143 & 100 & -6 &  112 & 100 & 0 \\
0.07  & 143 & 98  & -6 &  112 & 99 & 0 \\
0.05  & 137 & 96  & -- & 112 & 98 & -- \\
0.03  & 143 & 66  & 12 & 112 & 91 & 0 \\
0.02  & 137 & 41  & 18 & 112 & 74 & 0 \\
0.01  & 125 & 20  & 41 & 119 & 42 & 0 \\
\hline
0.10  & 108 & 99  & 0  & 155 & 77 & 14 \\
0.07  & 108 & 96  & 6  & 148 & 61 & 7 \\
0.05  & 108 & 71  & -- & 155 & 50 & -- \\
0.03  & 114 & 40  & 6  & 183  & 28  & -7\\
0.02  & 114 & 25  & 12 & 211  & 17 & -14 \\
0.01  & 120 & 11  & 41 & --  & --  & --\\
\hline
0.10 & 205  & 79 & 18  & 268 & 91 & 14\\
0.07 & 216  & 65 & 12  & 268 & 83 & 0\\
0.05 & 176  & 52 & --  & 282 & 75 & --\\
0.03 & 170  & 41 & -24 & 275 & 59 & -35\\
0.02 & 114  & 31 & -41 & 268 & 42 & -35\\
0.01 &  80  & 16 & -59 & 247 & 23 & -28\\
\hline
\hline
\multicolumn{2}{c}{N$_{tot}$ 1\,/\,2\,/\,3} & \multicolumn{2}{c}{649\,/\,321\,/\,153} &  &  \multicolumn{2}{c}{434\,/\,263\,/\,135}  \\
\hline
\end{tabular}
\end{center}
\end{table} 

In spot  3, the magnetic neutral line lies inside the umbra,  
because the spot is close to the limb.
Therefore, inside the sunspot the longitudinal component of the magnetic field 
along the line of sight is weaker than the transverse component.
Hence, the Stokes $V$ profiles are not reliable for the calibration purpose. 
For this reason, we use the linear polarization components instead. 
We decided to use the profile center of the Stokes $Q$ 
(position of the $\pi$ component) in this case and use the amplitude asymmetry of the 
$\sigma$ components in $Q$ as a measure of the line antisymmetry. 
The dispersion of the velocity reference points in the selected umbral pixels (spot 3) 
is slightly higher than for spots  1 and  2 (Table 2). 
We estimate a maximum calibration error for 
the spot 3 of $\sim$\,250 m\,s$^{-1}$. This does not affect our conclusions, because 
velocity values in spot 3 are larger than in spots 1 and 2. 
From the velocity dispersions in each class ($d_1$ and $d_2$), we estimate that 
we achieve a precision of $\pm$\,150 m\,s$^{-1}$ for the spots 1 and 2 
and $\pm$\,250 m\,s$^{-1}$ for the spot 3 in the TIP and POLIS data, 
respectively\footnote{Our convention is that positive values correspond to redshift.}.

\subsubsection{Absolute velocity calibration}
Using the telluric lines in the POLIS data, 
it is possible to perform an absolute velocity calibration 
for these data sets 
(Schmidt \& Balthasar 1994; Mart\'inez Pillet et al. 1997; Sigwarth et al. 1999). 
We use the 630.20\,nm line, because it is less influenced by 
the solar iron lines in the umbra than the O$_2$--line at 630.27\,nm. 
We only correct line--core positions for the curvature of the 
spectrograph as explained in Sect. 3.1. Then, the radial velocity 
between the observatory and the sunspot is determined. 
We use the values of Balthasar et al. (1986) for the sunspot angular rotation ([14.551,-2.87], cf.~their Eq.~(1)) to  
compute the radial velocity components: the earth rotation, the earth orbital motion 
and the solar rotation. 
Total radial velocities for the spots  1,  2, and  3 (sum of the three components) 
are 0.253, 0.862 and 1.327 km\,s$^{-1}$, respectively. 
These values take into account the line of sight component of the solar rotation, 
which amounts to 0.918, 1.438, and 1.971 km\,s$^{-1}$. 
Gravitational correction is also considered  for all 
solar lines (Table 3). Here, the rest--frame for each resolution element is 
the position of the telluric line plus a constant shift which includes 
proper correction for the radial velocity. 
In other words, there is a reference point in each resolution element different 
from the others. 

\begin{table}
\begin{center}
\caption{Wavelengths of the spectral lines in the POLIS data. 
The laboratory wavelengths are from Higgs (1960, 1962). 
The rest value includes the gravitational redshift 
($\Delta \lambda = 2.12 \times 10^{-6} \lambda$).}
\begin{tabular}{c c c c} 
\hline
 & Laboratory (\AA) & rest (\AA) & air (\AA)\\ \hline
Fe\,I  & 6301.4990   & 6301.5124  & - \\
O$_{2}$ & -         & -          & 6302.0005\\
Fe\,I & 6302.4920    & 6302.5054  & -\\
O$_{2}$ & -         & -          & 6302.7629\\
Ti\,I & 6303.750     & 6303.763   & -\\
\hline
\end{tabular}
\end{center}
\end{table}

\paragraph{Source of errors} 
Apart from the precision of the 
solar center and sunspot positions, the main challenge is to determine  
the exact sunspot angular rotation. 
Angular rotation of sunspots depends on the age and probably 
the phase of the solar cycle. 
It is also affected by the sunspot proper motion. So,  
the radial velocity values remain uncertain within $\sim$\,$\pm$\,100\,m\,s$^{-1}$. 
Moreover, there are some errors in positioning 
during the observations. 
For example, 5 arcsec error in the solar longitude creates an error of  
$\sim$\,15\,m\,s$^{-1}$ in the radial velocity (depending on the position). 
Pixel by pixel variation due 
to different time and slit positions cause a maximum error of  
$\sim$\,10\,m\,s$^{-1}$ between the first and the last scan steps.
Therefore, we conclude that for the absolute velocity calibration  
the precision is $\pm$\,150 m\,s$^{-1}$. 

Oscillations in the photospheric level of the umbra may induce  
errors in the Stokes $V$ calibration. 
However observations imply that such fluctuations are small. 
Mart\'inez Pillet et al. (1997), for example, 
studied a large sample of active regions and reported an umbral velocity 
of 65\,m\,s$^{-1}$ due to umbral oscillations. 
Balthasar \& Schmidt (1994) also investigated five minute 
oscillations in the umbra and reported an amplitude of 90\,m\,s$^{-1}$\,(rms).

\subsubsection{Comparison of the two calibration methods}
In order to check the assumption that the umbra is at rest, 
we compare the two calibration methods. 
Subtracting the corresponding velocity maps for Fe\,I 630.25 nm 
yields a difference map that show low amplitude variations, 
definitely less than the calibration uncertainties. It is important to note that we  
used a constant velocity reference point in 
the first method for the whole image, while in the absolute calibration    
we used different reference points (one for each spatial pixel).
To compare the absolute calibration with the first method, we define a 
mean reference point for the absolute calibration, 
the corrected telluric line position in the umbra.

\begin{table}
\begin{center}
\caption{Mean values and dispersions of the umbral velocities for the 630.25\,nm 
iron line in km\,s$^{-1}$. 
The last column shows uncertainties in the Stokes $V$ and absolute 
calibrations for the spots 1 and 2. These values for the spot 3 are 
0.211 and 0.282\,km\,s$^{-1}$ respectively. 
It confirms our principal assumption that the umbra is at rest 
(within our calibration error).} 
\begin{tabular}{c c c c} 
\hline
spot  1     & Stokes $I$ & Stokes $V$ & error\\ \hline
Stokes $V$ calib  &   0.085 $\pm$ 0.242 &   -0.023 $\pm$ 0.059 & 0.112 \\\hline
absolute calib  &       -0.135 $\pm$   0.258 &   -0.258 $\pm$ 0.060 & 0.176 \\ \hline 
spot  2   & & \\ \hline 
Stokes $V$ calib  &      -0.158 $\pm$ 0.143 &  -0.038 $\pm$ 0.088 & 0.176 \\ \hline
absolute calib  &      -0.168 $\pm$ 0.135 &  -0.044 $\pm$ 0.072 & 0.211 \\ \hline
\end{tabular}
\end{center}
\end{table}

Comparing Stokes $V$ and absolute calibrations, there are 
small differences between them and also small dispersions in each one:  
Table 4 lists the mean umbral velocities for the {\it whole umbra} for the Fe\,I 630.25\,nm 
line obtained from both methods. 
The average Stokes $V$ umbral velocities with absolute calibration confirm that our basic assumption, 
taking the umbra at rest, is correct. 
Average line--core velocities are also less than our calibration errors. 
Their tendency for a blueshift is probably due to 
stray light contamination from the quiet sun. 
However, in the Stokes $V$ absolute calibration there is a tendency for a blueshift 
because the average velocities of the whole umbra 
include umbral dots and periphery points. 
These lead to the weak blueshifts in the umbra, comparable to our calibration precision. 
There are reports of strong upflows in the umbra of preceding spots in  
bipolar pairs (e.g. Sigwarth 2000); however all three spots were isolated. 
Our results do not exclude that there are some oscillations 
in the umbra, but the amplitudes should be less than 
our precision limit of $\pm$\,150\,m\,s$^{-1}$ 
(as reported by, e.g. Lites et al. 1998; Bellot Rubio et al. 2000a). 

The absolute calibration confirms our first method based on the Stokes $V$ profiles of the umbra. 
Therefore, we can confidently use this method 
where an absolute calibration is not possible like for the TIP data.

\begin{figure*}
\centerline{\resizebox{18cm}{!}{\includegraphics*{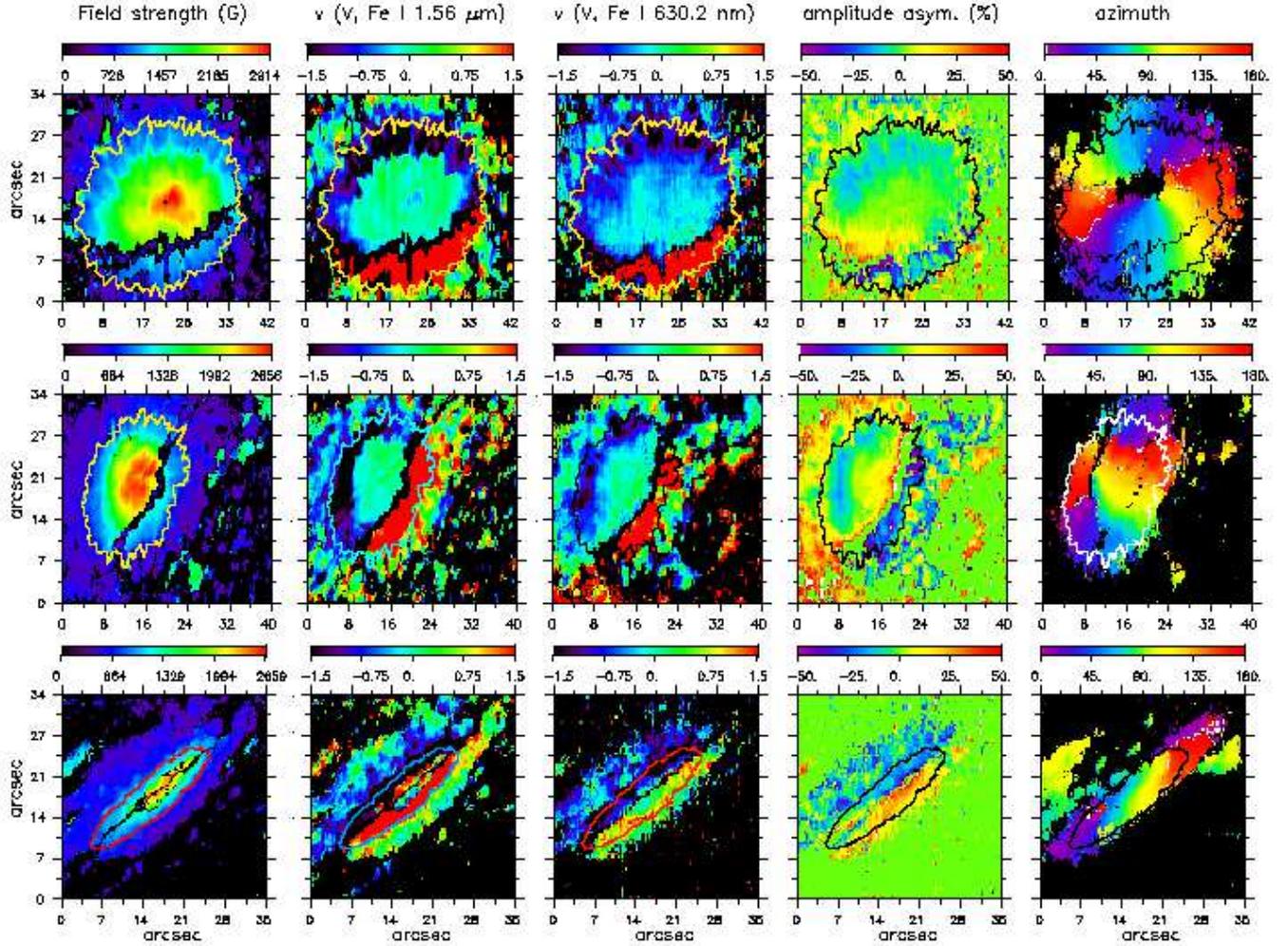}}}

\caption[]{{\em From left to right:} field strength, $B$, in Gauss ($10^{-4}$\,T); 
Stokes $V$ velocity in 1564.8\,nm in km\,s$^{-1}$; 
Stokes $V$ velocity in 630.25\,nm in km\,s$^{-1}$; Stokes $V$ amplitude asymmetry in 630.25\,nm (\%); 
 magnetic field azimuth in degree. 
{\em From top to bottom:} spot  1, spot  2, and spot  3. 
As it is clear in the Stokes $V$ amplitude asymmetry of spot  3, there are significant 
amplitude asymmetries in the umbra due to the presence of the magnetic neutral line. 
 Also note the usually larger velocities outside the penumbral boundary in the visible (630.25\,nm) 
than infrared (1564.8\,nm) line. 
Magnetic field strength and azimuth obviously indicate that the magnetic field of 
sunspots continue outside the white--light boundary with the same orientation. 
Separate magnetic elements may be easily distinguished in these maps.}
\end{figure*}


\subsection{Magnetic field parameters}
To investigate the flow field in the canopy, 
we calculate line parameters for all {\it regular} $V(\lambda)$ profiles, 
i.e. profiles which show two clear lobes. 
The quantities derived are positions and amplitudes of the peaks, 
area and amplitude asymmetries, and the magnetic field strength.  
We also analyze $Q(\lambda)$, $U(\lambda)$, and $L(\lambda)$ of the TIP data in order to have 
a measure for the inclination and azimuth of the magnetic field. 
We construct maps of field strength, azimuth, amplitude asymmetry of Stokes $V$, 
and Stokes $V$ velocities (Fig. 2). 
Comparing the second and third columns of Fig. 2, it is clear that 
the magnetic neutral line in the visible is shifted outward 
with respect to the infrared line.
The field strength is calculated using the Stokes $V$ splitting of the 1564.8\,nm line. 
The line is a Zeeman triplet which is in the strong field limit for $B >$ 0.03 T; 
so we can use the distance between the $\sigma$ component and the profile center 
as a measure of the magnetic field strength (e.g. Stix 2002) by:
$$ \Delta \lambda = 4.67 \times 10^{-18} \lambda^2 g_{\textrm{eff}} B$$
where $g_{\textrm{eff}}$ is the effective Land\'e factor, 
$\Delta \lambda$ and $\lambda$ are in nm, and 
$B$ is in tesla. This is not an exact method, especially for weak fields. 
So these estimates are upper limits for the weak magnetic field  of the canopy.
This method fails along the magnetic neutral line. 
Maps on the left column of Fig. 2 show the field strength.

Considering the fact that these Stokes profiles are formed in the weak field limit in the canopy, 
one may use the ratio of the $\sigma$ components of the linear and circular polarization 
to approximate the magnetic field inclination and azimuth (e.g. Stix 2002; Solanki 1993)
$$ \frac{V}{\sqrt{Q^2+ U^2}}\sim \frac{\cos \gamma}{\sin^2 \gamma} \ \ \ \textrm{ and} \ \ \ \ \frac{U}{Q}\sim \tan(2\chi) $$
where $\gamma$ is the inclination with respect to the line of sight and 
$\chi$ is the azimuth with respect to the celestial N--S direction.  
In the weak field limit, these ratios are independent of the field strength. 
Computing field orientation  
in this way is not as accurate as an inversion. 
However, it is sufficient for our purpose to check whether or not the field geometry 
in the canopy follows the same pattern as in the spot. 
The azimuth maps in Fig. 2 (right column) show 
the same pattern inside and outside the spot in the canopy. 
Thus, the canopy can be distinguished in these maps from the separate 
magnetic elements, which show different azimuth and field strength.

Because linear polarization signals are usually weaker than the circular polarization, 
it is not possible to compute the field azimuth for all pixels with a reasonable $V$ signal. 
Moreover, the field strength in the canopy is much smaller than in the spot. 
For this reason, we have smaller coverage in the azimuth maps, which come from linear polarization 
signals, with respect to other maps which are based on $V$ signals (Fig. 2). 
Pattern of various parameters 
outside the spot in Fig. 2 show a smooth change at the spot boundary. 
In other words, the magnetic field strength and orientation, and the flow field 
do not change {\it abruptly} at the sunspot white--light boundary.


\section{Radial outflow in the canopy}
In this section, we define the magnetic canopy to be distinguished from the 
isolated magnetic elements. We investigate Stokes $V$ profiles in radial cuts, both 
on the limb and center sides. 
Box--car averaging in radial directions is another method to investigate the canopy. 
Finally, we study the canopy extension in our sunspots.

\begin{figure*}
\resizebox{\hsize}{!}{\includegraphics{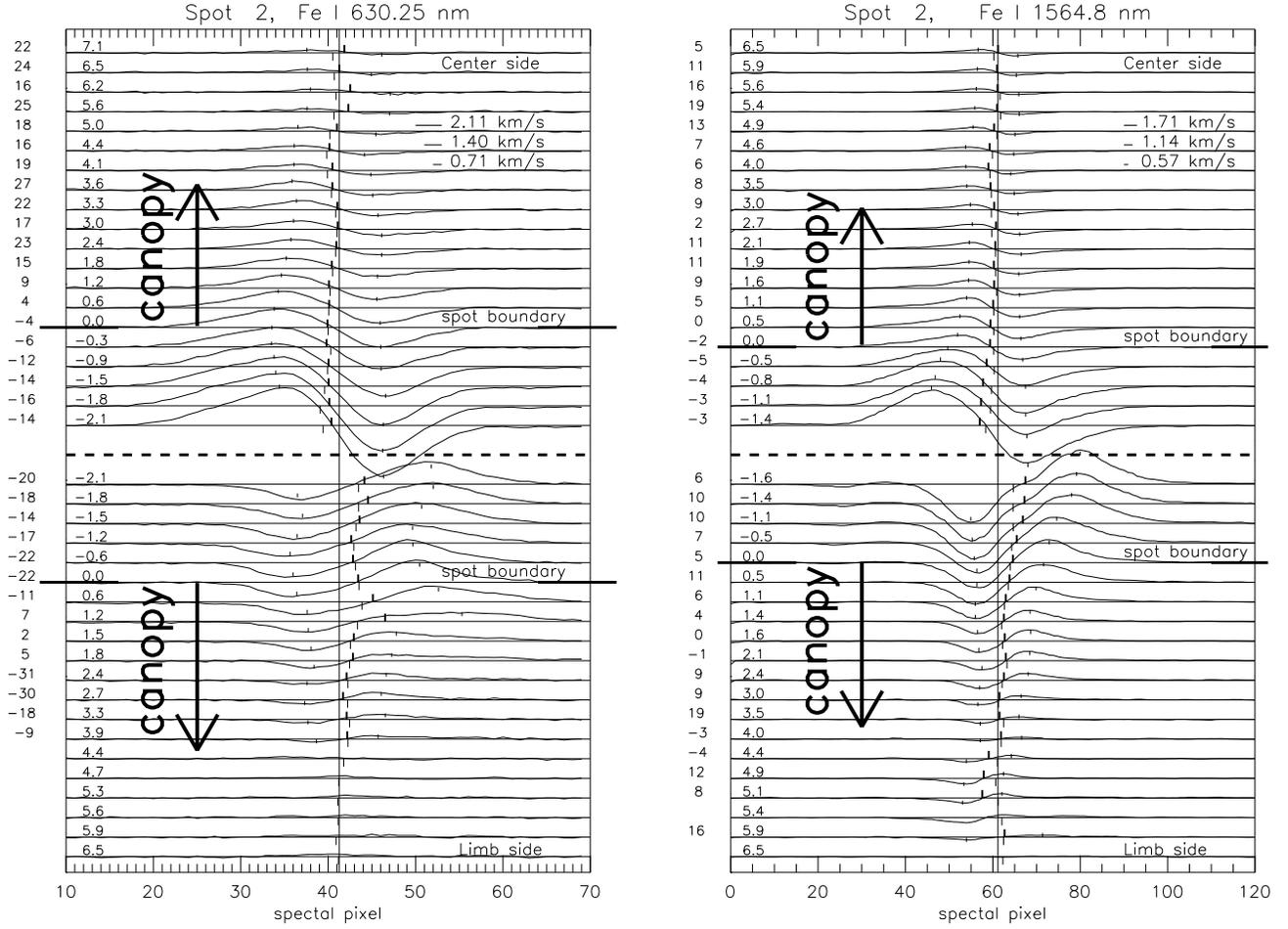}}
\caption[]{$V(\lambda)$ profiles in a radial cut in the axis of symmetry of 
 spot 2 in the POLIS (left) and TIP (right) data. 
Thin tick marks are positions of the max/min of the two lobes.
Thick tick marks above the horizontal line are the profile centers of Stokes $V$. 
Thin tick marks below the horizontal line show the corresponding Stokes $I$ (line--core) velocity. 
Vertical solid line is the velocity reference point based on the umbral calibration.
Numbers on the left column inside the diagram are distance of each profile  
from the spot boundary in arcsec. 
The left hand side numbers outside the diagram are the amplitude asymmetries in percent. 
Note the usually larger amplitude asymmetries in the visible data. 
The thick dashed horizontal lines 
show the discontinuity in the diagram and separate the limb and center side. 
Three small lines indicate the amounts of required shifts for some velocities. 
Sunspot boundaries in the limb and center side are indicated. }
\end{figure*}

\subsection{Canopy definition and isolated magnetic elements}
The maximum distance at which there is a magnetic field along with a 
flow (with consistent {\it sign and gradient}) varies for different spots. 
We attribute a pixel in a radial cut to the sunspot canopy, if it demonstrates 
all the properties below:
\begin{enumerate}

\item It has the same magnetic polarity and velocity sign as the 
preceding pixel closer to the spot boundary.

\item It shows the same inclination and azimuth patterns as the 
preceding pixel closer to the spot boundary.

\item Both the absolute values of the velocity and magnetic field strength 
decrease with respect to the preceding pixel or remain constant.

\end{enumerate}
The right column of Fig. 1 illustrates the polarity maps. 
It is clear that close to each spot, there is 
a unipolar--connected region that we attribute to the canopy. 
But as we go outward, some isolated magnetic elements 
appear with either different polarities or velocity signs. 
Therefore, the {\it sign and gradient} of velocity at some of these 
points do not show a monotonic behavior. 
These isolated magnetic elements are seen in various maps of 
Figs. 1, 2 (e.g. at $x = 36$ and $y = 24$ arcsec in the spot 2).

\subsection{Stokes $V$ profiles in radial cuts}
Fig. 3 shows Stokes $V$ profiles in a radial cut 
in the limb and center side directions (axis of symmetry)  
for the spot 2 in the TIP and POLIS data.  
In this  figure, the Stokes $V$ velocity smoothly decreases outward 
{\it without any abrupt change} at the sunspot boundary. 
The sign and gradient of the velocities remain constant at first, but start to change 
in the outer parts. 
Note the very broad red lobes in the limb side in the visible data, 
while the corresponding profiles in the infrared data are not so much affected 
(but still are broader than the blue lobes).  
It may be due to the presence of magnetic structures high in the atmosphere 
such that they cannot significantly affect the 1.56 $\mu$m lines. 
Stokes $V$ velocities of these broad profiles have to be considered with caution.
Around 4 arcsec from the penumbral boundary in the center side, 
there is a jump in velocity in both  lines. Further out on this side, 
the velocity changes sign. This indicates that these points cannot 
be attributed to the sunspot canopy. 
Another interesting feature is a separate magnetic element $\sim$\,5 arcsec 
from the spot boundary on the limb side in the infrared line. 
Within two arcsec, redshift of the profile center converts to blueshift and then 
changes to redshift. Also note the amplitude of the $V(\lambda)$ profile, which 
decreases and then increases. 
The same is true for the magnetic field strength (Stokes $V$ splitting). 
The amplitude asymmetry of each profile is also indicated in Fig. 3. 
There are large amplitude asymmetries 
which may be due to gradients or jumps in the magnetic field and/or velocity (e.g. Bellot Rubio et al. 2000b).

\subsection{Box--car averages of the flow and field parameters in the sunspot symmetry axis}

Comparison of the line--core and Stokes $V$ velocities plays an important role 
in distinguishing the radial outflow in the canopy and the moat flow. 
We average velocities in small square boxes along the symmetry axis of the sunspots. 
We have large coverage outside the spot in the limb side of the spot 2;  
so we study this case in detail.  
Inside the penumbra, the line--core velocities are smaller than   
the Stokes $V$ velocities, having a similar radial dependence (cf. Fig. 4). 
Outside the spot, the curves of line--core and Stokes $V$ velocities intersect each other. 
Thus, a few arcsec outside the penumbral boundary, the 
value of the Stokes $V$ velocity in the visible/infrared   
is less than the corresponding Stokes $I$ velocity. 
The intersection points for the infrared lines are slightly 
closer to the sunspots than visible lines.
The range of the velocities in the canopy is $\sim$\,0.5\,--\,2.0 km\,s$^{-1}$. 
High velocity patches are rare. Roughly speaking, 
a velocity of 1.0 km\,s$^{-1}$ is common in the canopy of all the three spots. 
This is far from the average values for the line--core velocities. These are around 
0.6\,--\,0.7 km\,s$^{-1}$ close to the sunspots,  
reaching  $\sim$\,--\,0.2 km\,s$^{-1}$ 
(the typical values for the convective blueshift) at the boundary of the moat cell. 
At these distances, there is no uniformly connected canopy as described above. 
Magnetic fields of these regions are governed by the plage activity 
and isolated magnetic elements floating in the sunspot moat.

\begin{figure}
\resizebox{\hsize}{!}{\includegraphics{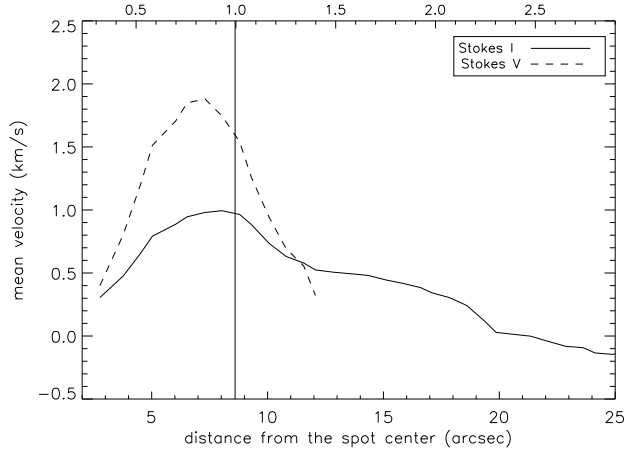}}
\caption[]{The solid line is the line--core velocity and the dashed line 
is the Stokes $V$ velocity of Fe\,I 1564.8\,nm in the limb side of the spot 2. 
The vertical line is the penumbral boundary 
($\pm$\,0.2 arcsec for differences in neighbor pixels). 
Inside the spot, $V$ and $I$ Stokes velocities follow each other,  
while they show different behaviors outside the spot. 
Upper x-axis is the distance from the penumbral boundary in units of 
sunspot radius. The averaging box is 17$\times$17 
pixels to cover granules and intergranular lanes.\label{fig5}}
\end{figure}

\begin{figure}
\resizebox{\hsize}{!}{\includegraphics{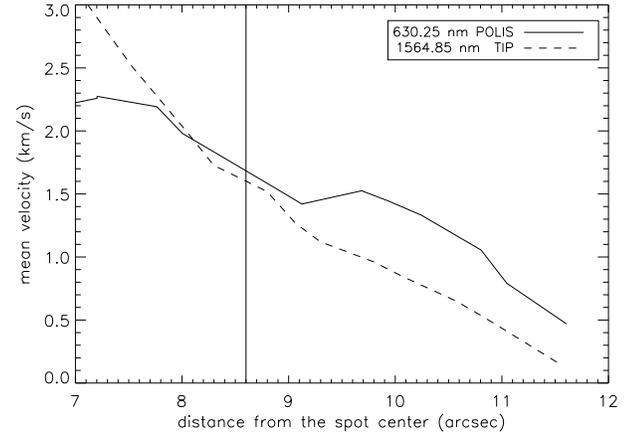}}
\caption[]{This plot shows the radial variation on Stokes $V$ velocities on the limb side 
of the spot 2. 
The solid line is the visible Fe\,I 630.25\,nm Stokes $V$ velocity and the dashed line 
is the same for the near IR Fe\,I 1564.8\,nm. 
The vertical line is the penumbral boundary ($\pm$\,0.2 arcsec for differences in neighbor pixels). 
Inside the sunspot, the TIP data show larger velocities while outside the spot, 
the POLIS data show larger values. The averaging box is 5$\times$5 pixels in this diagram.}
\end{figure}

\begin{figure}
\resizebox{\hsize}{!}{\includegraphics{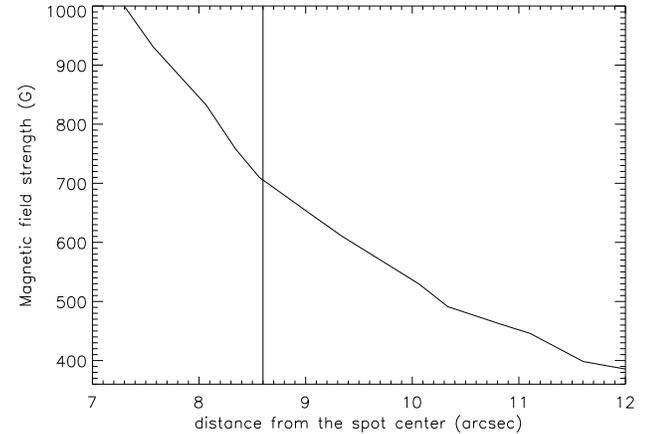}}
\caption[]{Same as Fig. 5 for the magnetic field strength in the TIP data.}
\end{figure}

The contribution function of the iron lines at 630\,nm peaks in higher layers relative to  
the lines at 1.56 $\mu$m. 
Visible lines form in larger geometrical extents than the infrared lines. 
In addition, Stokes $V$ profiles form exclusively in the magnetic part of the atmosphere. 
The fact that we usually observe larger Stokes $V$ 
velocities in the visible than in the infrared in the canopy (Figs. 2,\,5)  
implies that there is a flow in the magnetic part of the atmosphere which affects the 
visible lines more than the infrared lines.
Observations by Balthasar et al. (1996) of the moat flow in non--magnetic lines 
resulted in low amplitude smooth velocity field ($\sim$\, 0.5 km\,s$^{-1}$), 
similar to velocities we find in the Stokes $I$ data.  
Beside this, the spatial extension of the moat flow is much more than 
what we observe in the Stokes $V$ maps. 
There are also significant amplitude asymmetries in the sunspot canopy (Figs. 2,\,3) 
which indicate that there are gradients or jumps in the velocity field. 
Therefore, the line--core velocity is comparable with the moat flow velocity in non--magnetic lines 
while the Stokes $V$ velocity is not a reasonable proxy for the moat flow.
Hence, we attribute line--core velocity to the moat flow and Stokes $V$ velocity to the canopy outflow.

We find average amplitude asymmetries in the canopy profiles of $\sim$\,10\,\%. 
These values are less than the maximum amplitude asymmetry, which 
peaks in the middle/outer penumbra (Schlichenmaier \& Collados 2002). 
Amplitude asymmetries of the canopy in spot 2 are given in Fig. 3. 
The profile asymmetry is one of the typical properties 
of the Evershed flow, which we observe in the canopy radial outflow. 
It is different from the amplitude asymmetry in plage and network 
which changes from one pixel to the next one. 
We usually observe gradually changing amplitude asymmetries 
in nearby pixels (Fig. 3).

The magnetic field strength decreases smoothly outside the spot (Fig. 6). There is no sharp 
change at the penumbral boundary or within 1 arcsec. 
The lobe separation in the Stokes $V$ profiles of the canopy is a measure 
of the minimum magnetic field. The smallest Stokes $V$ splitting observed are $\sim$ 0.03 T. 
As we go outward and 
leave the uniform--connected canopy with coherent flow field, disturbances in the observed 
$V(\lambda)$ profiles increase. 
At some point, Stokes $V$ splitting increases while the amplitude 
decreases (Fig. 3, limb side, Fe\,I 1564.8\,nm line, profile at 5.9 arcsec).

The average canopy extensions of 1.2, 1.35, and 1.6 ($\pm$ 0.05) penumbral 
radii for the spots 1, 2, and 3  depend almost linearly on the cosine of the heliocentric angle.
We observe similar canopy extensions in the visible and infrared lines. 
All the three sunspots were observed with the same setup and exposure time. 
It may be possible to trace larger canopies close to the disk center 
with longer exposure times, and hence higher signal-to-noise ratios.

\section{Conclusion}

We investigate the flow and magnetic field in the immediate surroundings of three sunspots 
using spectropolarimetric data observed simultaneously at the VTT in two spectral 
regions around Fe\,I\,1564.8\,nm and around Fe\,I\,630.2\,nm, with TIP and POLIS, respectively. 
The existence of a radial outflow in the sunspot canopy is a matter of strong debate. 
Here, we provide strong evidence for the existence of the canopy, and for the existence 
of a radial outflow in the canopy by analyzing Stokes profiles. 

To this end it is essential to properly calibrate the velocity. We use two different 
calibration methods. The first one is based on the assumption that the most antisymmetric 
profiles in the umbra reflect locations at rest. The second is an absolute calibration 
which uses the telluric line in the vicinity of Fe\,I\,630.2\,nm, taking into account 
the relative motions between the telescope and the surface of the sun. 
Both methods are consistent, which implies that antisymmetric $V$ profiles in the 
umbra represent a rest--frame.

Evidence for the existence of a canopy surrounding all three sunspots at various 
heliocentric angles is presented:\\
(1) The polarization signal extends outside the sunspot white--light boundary 
up to 1.2, 1.35, and 1.6 penumbral radii for spots at $\mu$\,=\,0.89, 
0.64, and 0.26 respectively.\\
(2) We find that  velocities and the estimated magnetic field strength 
vary smoothly across the penumbral boundary. 
There is no abrupt change within one arcsec, 
which corresponds to our spatial resolution.\\
(3) The field azimuth and polarity in the canopy follow the spot pattern.\\ 
(4) Velocities determined from Doppler shifts of Stokes $V$ show a radial outflow 
in the surroundings of sunspots. The same is true for line--core velocities 
from Stokes I, but as argued in Sect. 4.3, the latter is related to the moat flow.\\ 
(5) Significant amplitude asymmetries of Stokes $V$ exist almost everywhere in 
the surroundings of sunspots, indicating depth 
gradients or discontinuities in the flow velocity. 

All our findings are consistent with a magnetic canopy surrounding sunspots, 
which exhibits a discontinuity (magnetopause) rising with radial distance from 
the spot. Moreover, our findings indicate that the radial outflow in the canopy 
is an extension of the penumbral Evershed flow.
The decreasing $V$--velocity with radial distance from the spot is consistent 
with a rising canopy base, i.e. with a magnetopause which rises outwards from 
the spot. Line--core velocities of Stokes $I$ persist outwards in the surroundings 
of sunspots and are therefore attributed to the moat flow. As one would expect, 
the canopy extension increases with heliocentric angle, and even at a small 
heliocentric angle of 27$^\circ$ the canopy extension is at least 3 arcsec, 
much more than the spatial resolution. 
Therefore, we conclude that at least a part of the 
penumbral Evershed flow continues in a magnetic canopy that surrounds sunspots.

\begin{acknowledgements}
We wish to thank Wolfgang Schmidt, Hubertus W\"ohl, and Dirk Soltau for useful discussions. 
Daniel Cabrera Solana (IAA) kindly provided the routine for the radial velocity calculation. 
\end{acknowledgements} 


\begin{thebibliography}{}

\bibitem[1988]{Alissandrakis} Alissandrakis, C.E., Dialetis, D., Mein, P. et al., 1988, A\&A 201, 339

\bibitem[1996]{Ballesteros} Ballesteros, E., Collados, M. \& Bonet, J.A., 1996, A\&AS 115, 353

\bibitem[1994]{Balthasar1} Balthasar, H. \&  Schmidt, W.,  1994, A\&A 290, 649

\bibitem[1986]{Balthasar2} Balthasar, H., V\'azquez, M. \&  W\"ohl, H.,  1986, A\&A 155, 87

\bibitem[1996]{Balthasar3} Balthasar, H., Schleicher, H., Bendlin, C. \& Volkmer, R., 1996, A\&A 315, 603

\bibitem[2005]{Beck1} Beck, C., Schmidt, W., Kentischer, T. \& Elmore, D., 2005a, A\&A 437, 1159

\bibitem[2005]{Beck2} Beck, C., Schlichenmaier, R., Bellot Rubio, L.R. \& Kentischer, T., 2005b, 
A\&A, 443, 1047

\bibitem[2004]{Bellot Rubio1} Bellot Rubio, L.R., 2004, Sunspots as seen in polarized light, 
Reviews in Modern Astronomy, 17, Ed. R.E. Scielicke, Wiley

\bibitem[2000a]{Bellot Rubio2} Bellot Rubio, L.R., Collados, M., Ruiz Cobo, B. et al., 2000a, ApJ 534, 989

\bibitem[2000b]{Bellot Rubio3} Bellot Rubio, L.R., Ruiz Cobo, B. \& Collados, M.,  2000b, ApJ 535, 489

\bibitem[2003]{Bellot Rubio4} 
Bellot Rubio, L.R., Balthasar, H., Collados, M. \& Schlichenmaier, R., 2003, A\&A 403, L47

\bibitem[1992]{Borner} B\"orner, P. \& Kneer, F., 1992, A\&A 255, 307

\bibitem[1963]{Brekke} Brekke, K. \& Maltby, P., 1963, Ann. d'Astrophys. 26, 383

\bibitem[2005]{Cabrera Solana} Cabrera Solana, D.,  Bellot Rubio, L.R., 
\& del Toro Iniesta, J.C., 2005, A\&A 439, 687

\bibitem[1999]{Collados} Collados, M., 1999, in ASP Conf. Ser. 184, Thirds Advances in Solar
 Physics Euroconference: Magnetic Field and Oscillation, 3
 
\bibitem[1985]{Dialetis} Dialetis, D., Mein, P., \&  Alissandrakis, C.E., 1985, A\&A 147, 93

\bibitem[1909]{Evershed} Evershed, J., 1909, MNRAS, 69, 454
 
\bibitem[1982]{Giovanelli} Giovanelli, R.G. \& Jones, H.P., 1982, Solar Phys 79, 267

\bibitem[1960]{Higgs1} Higgs, L.A., 1960, MNRAS, 121, 421

\bibitem[1962]{Higgs2} Higgs, L.A., 1962, MNRAS, 124, 51

\bibitem[2001]{Hirzberger} Hirzberger, J. \& Kneer, F., 2001 A\&A 378, 1078

\bibitem[1985]{Kuveler} K\"uveler, G. \& Wiehr, E.,  1985, A\&A 142, 205

\bibitem[1983]{Landi DeglInnocenti} 
Landi Degl'Innocenti, E. \& Landolfi, M., 1983, Solar Phys 87, 221

\bibitem[2001]{Leka} Leka, K.D., \&  Steiner, O., 2001, ApJ 552, 354

\bibitem[1998]{Lites} Lites, B.W., Thomas, J.H., Bogdan, T.J. et al., 1998, ApJ 497, 464

\bibitem[1997]{Martinez Pillet1} Mart\'inez Pillet, V., Lites, W. \&  Skumanich, A., 1997, ApJ 474, 810
 
\bibitem[1999]{Martinez Pillet2} Mart\'inez Pillet, V., Collados, M., Sanchez Almeida, J. et al., 1999,
in ASP Conf. Ser. 183, High Resolution Solar Physics: Theory, 
Observation, and Techniques, 264

\bibitem[1995]{Rimmele1} Rimmele, R.T., 1995a, ApJ 445, 511

\bibitem[1995]{Rimmele2} Rimmele, R.T., 1995b, A\&A 298, 260

\bibitem[2002]{Schlichenmaier1} Schlichenmaier, R. \& Collados, M., 2002, A\&A 381, 668

\bibitem[2000]{Schlichenmaier2} Schlichenmaier, R. \& Schmidt, W., 2000, A\&A 358, 1122

\bibitem[1994]{Schmidt1} Schmidt, W. \& Balthasar, H., 1994, A\&A 283, 241

\bibitem[1995]{Schmidt2} Schmidt, W. \& Kentischer, T., 1995, A\&AS 113, 363

\bibitem[2003]{Schmidt3} Schmidt, W., Beck, C., Kentischer, T. et al., 2003, AN 324, 300

\bibitem[1989]{Schroter} Schr\"oter, E.H., Kentischer, T., \& M\"unzer, H., 1989, in High Spatial 
Resolution Solar Observations, O. van der L\"uhe (Ed.), National Solar Obs., 
Sunspot, NM, p. 299

\bibitem[1972]{Sheeley} Sheeley N.R., Jr., 1972, Solar Phys 25, 98

\bibitem[2000]{Sigwarth1} Sigwarth, M., 2000, Rev Mod Ast, 13, 45

\bibitem[1999]{Sigwarth2} Sigwarth, M., Balasubramaniam, K.S., Kn\"olker, M. 
\& Schmidt, W., 1999, A\&A 349, 941

\bibitem[1993]{Solanki1} Solanki, S.K., 1993, Space Sci Rev, 63, 1

\bibitem[2003]{Solanki2} Solanki, S.K., 2003, A\&AR 11, 153

\bibitem[1994]{Solanki3} Solanki, S.K., Montavon, C.A.P., \& Livingston, W., 1994, A\&A 283, 221

\bibitem[1999]{Solanki4} Solanki, S.K., Finsterle, W., R\"uedi, I. \& Livingston, W., 1999, A\&A 347, L27

\bibitem[2000]{Steiner} Steiner, O., 2000, Solar Phys 196, 245

\bibitem[2002]{Stix} Stix, M., 2002, The Sun: An Introduction,  Springer

\bibitem[1993]{Title} Title, A.M., Frank, Z.A., Shine, R.A. et al., 1993, ApJ 403, 780

\bibitem[1996]{Wiehr1} Wiehr, E., 1996, A\&A  309, L4

\bibitem[1989]{Wiehr2} Wiehr, E. \& Balthasar, H., 1989, A\&A  208, 303

\bibitem[1986]{Wiehr3} Wiehr, E., Stellmacher, G., Kn\"olker, M. \&  Grosser, H., 1986, A\&A  155, 402

\end{thebibliography}
\end{document}